\documentclass[a4paper,12pt]{article}

\usepackage{subfigure}
\usepackage{graphicx}
\usepackage{floatflt}
\usepackage{wrapfig}

\begin{document}

\title{ \textsc{separability and entanglement \\ of four-mode gaussian
states}}

\author{\bf Vladimir I. Man'ko$^{1}$ and Alexandr A. Sergeevich$^{2}$}
\date{}
\maketitle

\centerline{$^{1}$\textit{P. N. Lebedev Physical Institute, Russian
Academy of Sciences,}}

\centerline{\textit{Leninskii Prospect 53, Moscow 119991, Russia}}

\centerline{$^{2}$\textit{Moscow Institute of Physics and Technology
(State University),}}

\centerline{\textit{Institutskii per. 9, Dolgoprudnyi, Moscow Region
141700, Russia}} \vspace{5mm} \centerline{e-mails:
manko@sci.lebedev.ru$\;\;$ a.sergeevich@gmail.com}

\def\thesection{\arabic{section}.}
\makeatletter
\renewcommand\@biblabel[1]{#1.}
\makeatother

\vspace{5mm}
\begin{abstract}
\noindent The known Peres-Horodecki criterion and scaling criterion
of separability are considered on examples of three-mode and
four-mode Gaussian states of electromagnetic field. It is shown that
the principal minors of the photon quadrature dispersion matrix are
sensitive to the change of scaling parameters. An empirical
observation has shown that the bigger the modulus of negative
principal minors, the more entangled the state.
\end{abstract}


\setcounter{section}{0} \setcounter{equation}{0}

\section{Introduction}

The entanglement phenomenon is one of the most important quantum
properties of multimode Gaussian states (both pure and mixed) was
studied in the context of their separability and entanglement in
\cite{Simon, Duan} where the Peres-Horodecki criterion of
separability \cite{Peres, Horodecki} was implemented for detecting
the entanglement of two-mode-light Gaussian state. The
Peres-Horodecki criterion of separability of a quantum state is
based on the property of positive but not completely positive map of
the density matrix \cite{Sudarshan}, which is the transpose
transform of the density matrix. For a separable state of a
bipartite system, the partial transpose transform applied to
variables of one subsystem yields a new density matrix that
corresponds to the physical state. For the density matrix of an
entangled state, the partial transpose transform can give a matrix
that is a negative Hermitian matrix. The criterion (called
ppt-criterion) was used in \cite{Simon, Duan} to study the
entanglement of Gaussian states of photons. Later on, other
researchers successfully applied the ppt-criterion (see, for
example, [6--9]).

The other separability criterion based on a nonpositive map of the
density matrix, called the partial scaling transform, was suggested
in \cite{MS1, MS2}; its implementation is connected with properties
of the Wigner function \cite{Wigner} with respect to the scaling of
momentum. For a separable state of a two-mode light, the scaling of
the second component of momentum provides an admissible Wigner
function. The scaling criterion can be extended to the case of
multipartite quantum states. The ppt-criterion is a special case of
the scaling criterion with the choice of the scaling parameter equal
to $-1$. The scaling criterion was used recently by Chirkin and
Saigin \cite{Chirkin1, Chirkin2} to study the entanglement
properties of three-mode light generated in nonlinear crystals due
to the interaction, where three different processes are involved,
namely, two parametric down-conversions and one parametric
up-conversion.

In this work, we study the application of the scaling criterion to
three- and four-mode photon states. We consider both the pure and
mixed states. Our aim is to find some empirical correlations of the
degree of entanglement and negative values of principal minors of
the photon quadrature dispersion matrix as functions of the scaling
parameters.

The paper is organized as follows.

In Sec. 2 the definition of separability is discussed. In Sec. 3 the
Peres-Horodecki criterion is reviewed. The properties of scaling
transform are discussed in Sec. 4. The application of the scaling
criterion to the three-mode case is elaborated in Sec.5, while the
four-mode case is studied for a pure Gaussian photon state in Sec.
6. A mixed four-mode Gaussian state is considered in Sec. 7. Some
conclusions and perspectives are presented in Sec. 8.

\section{Separability and Entanglement}

We consider a single-mode photon state described by the quadrature
operators $\hat q$ and $\hat p$. The density matrix $\rho$ of any
field state obeys the following conditions:
\begin{equation}\rho^+ = \rho,\;\;\;\mbox{Tr} \rho = 1,\;\;\;\rho\geq0.\end{equation}
When taking a transpose of the density matrix, these conditions are
kept for a new 'transposed' state:
\begin{equation}\left(\rho^T\right)^+ = \rho^T,\;\;\;\mbox{Tr} \rho^T = 1,\;\;\;\rho^T\geq0. \end{equation}
The transpose transform of the density matrix is equivalent to the
time inversion or complex conjugation of the density matrix. There
is a one-to-one correspondence between density operators and Wigner
quasidistribution functions $W(q,p)$. From the definition of the
Wigner function it follows that the transpose transform of $\rho$ is
equivalent to a mirror reflection of momentum $p$ in the phase
space, namely,
\begin{equation}\rho\longrightarrow\rho^T \;\Longleftrightarrow\;
W(q,p)\longrightarrow W(q,-p). \end{equation}

Now we consider a two-mode state with the density operator $\hat
\rho (1,2)$ and annihilation operators
$$\hat a_1 = \frac{\hat q_1 + i\,
\hat p_1}{\sqrt 2}\,,\;\;\hat a_2 = \frac{\hat q_2 + i \,\hat
p_2}{\sqrt 2}.$$ We say that a state is simply separable if
\begin{equation}\hat \rho (1,2) = \hat \rho (1) \otimes \hat \rho(2). \end{equation}
A separable two-mode state is defined as a state with the density
operator representable as a convex sum of simply separable states:
\begin{equation}\hat \rho (1,2) = \sum\limits_j {k_j \, \hat \rho ^{(j)} _1 (1)
\otimes \hat \rho ^{(j)} _2 (2)}, \;\;\;\;\;\; \sum\limits_j {k_j} =
1, \;\;\; k_j\geq0. \end{equation} Here $\hat \rho_1 ^{(j)} (1)$ and
$\hat \rho_2 ^{(j)} (2)$ are the density operators of modes 1 and 2,
respectively. The states which cannot be presented in such a form
are called entangled states. In other words, by definition, an
entangled state is a state which cannot be presented as a convex sum
of simply separable states.

\section{The Peres-Hrodecki Criterion}

It is evident from (4) that the partial transpose, i.e., the
transpose of a matrix for the second-mode term ${\rho^{(j)} _2 (2)
\rightarrow \left(\rho^{(j)} _2 (2)\right)^T}$, leads to a separable
positive density operator
\begin{equation}\hat \rho^{pt} (1,2) = \sum\limits_j {k_j \, \hat \rho
^{(j)} _1 (1) \otimes \left(\hat \rho ^{(j)} _2
(2)\right)^T}.\end{equation} Thus, $\hat \rho^{pt}(1,2)$ describes a
new separable state.

Summarizing, we can say that the partial transpose
$\hat\rho\rightarrow\hat\rho^{pt}$ of a separable state necessarily
gives the density operator, which fits the conditions (1). This is
the Peres-Horodecki criterion of separability.

Rewriting the decomposition (5) in terms of Wigner functions we
obtain
\begin{equation}W(q_1,p_1,q_2,p_2)=\sum\limits_j
{l_j\,W^{(j)} _1 (q_1,p_1)W^{(j)} _2 (q_2,p_2)}. \end{equation} Let
us consider the Wigner function of the generic Gaussian form
\begin{equation}W(q,p)=\frac{1}{\sqrt{\mathrm{det}\,
\sigma}}\exp\left(-\frac{1}{2} \bf Q \sigma^{-1} \bf Q ^T\right),
\end{equation}
where the four-dimensional vector $\bf Q$ reads
\begin{equation}\mathbf{Q}=\left(q_1-\left\langle q_1 \right\rangle, q_2-\left\langle
q_2 \right\rangle, p_1-\left\langle p_1 \right\rangle,
p_2-\left\langle p_2 \right\rangle \right) \end{equation} and the
matrix $\sigma$ is a $4\times4$ real symmetric variance matrix
\begin{equation}\sigma_{r_i r_j}=\frac{1}{2}\left\langle \hat r_i \hat r_j + \hat r_j \hat r_i \right\rangle, \end{equation}
with $\hat r_1 = \hat q_1, \; \hat r_2=\hat q_2,\; \hat r_3=\hat
p_1,\; \hat r_4=\hat p_2$.

From (3) and (8) it is easy to see that the partial transpose is
equivalent to the transform of a variance matrix $\sigma \rightarrow
\sigma'$, which is defined as follows:
\begin{equation}\left(\begin{array}{cccc}
\sigma_{q_1q_1} & \sigma_{q_1q_2} & \sigma_{q_1p_1} & \sigma_{q_1p_2} \\
\sigma_{q_2q_1} & \sigma_{q_2q_2} & \sigma_{q_2p_1} & \sigma_{q_2p_2} \\
\sigma_{p_1q_1} & \sigma_{p_1q_2} & \sigma_{p_1p_1} & \sigma_{p_1p_2} \\
\sigma_{p_2q_1} & \sigma_{p_2q_2} & \sigma_{p_2p_1} &
\sigma_{p_2p_2}
\end{array} \right)
\longrightarrow \left(\begin{array}{cccc}
\sigma_{q_1q_1} & \sigma_{q_1q_2} & \sigma_{q_1p_1} & -\sigma_{q_1p_2} \\
\sigma_{q_2q_1} & \sigma_{q_2q_2} & \sigma_{q_2p_1} & -\sigma_{q_2p_2} \\
\sigma_{p_1q_1} & \sigma_{p_1q_2} & \sigma_{p_1p_1} & -\sigma_{p_1p_2} \\
-\sigma_{p_2q_1} & -\sigma_{p_2q_2} & -\sigma_{p_2p_1} &
\sigma_{p_2p_2}
\end{array}\right).
\end{equation}
The conditions of existence of a partial transposed state in the
Gaussian case can be rewritten in the form of the
Robertson-Schr$\mathrm{\ddot o}$dinger uncertainty relation
\cite{Schrodinger, Robertson}. Thus, the Peres-Horodecki criterion
of separability is equivalent to the condition of positivity of
principal minors of the matrix
\begin{equation}\Sigma = \sigma' + \frac{i}{2}\Omega, \end{equation}
where
\begin{equation}\Omega = \left(\begin{array}{cccc}
0 & 0 & -1 & 0 \\
0 & 0 & 0 & -1 \\
1 & 0 & 0 & 0 \\
0 & 1 & 0 & 0
\end{array}\right). \end{equation}
Here the dimensionless units with $\hbar=1$ are used.

Since the transform considered does not change the minors but the
determinant, the Peres-Horodecki criterion can be simplified into
the inequality
\begin{equation}\det \Sigma \geq 0. \end{equation}

This criterion is a necessary and sufficient condition of
separability of Gaussian states in the two-mode case, but the
application of the Peres-Horodecki criterion to the $n$-mode case
with $n\geq 3$ fails to detect all entangled states.

\section{Scaling Transform}

Now we consider a scaling transform, which is applicable for
detecting the entanglement of Gaussian states. We generalize the
reflection transform $p\rightarrow -p$ in the phase space and use
the scaling of momentum $p\rightarrow \lambda p$. In that way, we
define the map
\begin{equation}\left \{ {\begin{array}{lcl} q\longrightarrow q, \\ p \longrightarrow \lambda p. \end{array}} \right. \end{equation}
For $\lambda \in [-1,1]$, this map defines a semigroup of maps. The
map (15) is equivalent to the scaling of time $t\rightarrow\lambda
t$. In the case  $\lambda=1$, this is an identical map; for
$\lambda=-1$ we have a partial transpose used in the Peres-Horodecki
criterion of separability, which was discussed in the previous
section.

Upon applying the scaling transform, the single-mode Wigner function
$W(q,p)$ modifies as follows:
\begin{equation}W(q,p)
\longrightarrow W_\lambda (q,p) = NW(q,\lambda p), \end{equation}
where $N$ is a normalization constant.

The initial Wigner function is normalized
\begin{equation}\int W(q,p)\frac{dq\,dp}{2\pi}=1, \end{equation}
and the transformed Wigner function must be normalized, as well:
\begin{equation}\int
W_\lambda(q,p)\frac{dq\,dp}{2\pi}=\int N W(q,\lambda
p)\frac{dq\,dp}{2\pi}=1. \end{equation} Substituting $p'=\lambda p$
into the integral (18), we obtain
\begin{equation}\int NW(q,p')\frac{dq\,dp'}{2\pi|\lambda|}=1. \end{equation}
Taking into account (16) we have the following result for the
normalization constant $N$:
\begin{equation}\frac{N}{|\lambda|}=1\; \Longrightarrow \;N=|\lambda|. \end{equation}
Hence, the scaling transform modifies the Wigner function as
follows:
\begin{equation}W(q,p)\longrightarrow W_\lambda (q,p) = |\lambda| W(q,\lambda p). \end{equation}

Now we find out how the partial scaling transform affects the
dispersion matrix of the two-mode quantum state. We perform the
scaling of the second momentum by the factor of $\lambda$,
\begin{equation}p_2
\longrightarrow \lambda p_2. \end{equation}
Obviously, the variance
matrix $\sigma$ converts into
\begin{equation}\sigma_\lambda = \left(\begin{array}{cccc}
\sigma_{q_1q_1} & \sigma_{q_1q_2} & \sigma_{q_1p_1} &
\frac{1}{\lambda}\sigma_{q_1p_2}
\\ \\
\sigma_{q_2q_1} & \sigma_{q_2q_2} & \sigma_{q_2p_1} &
\frac{1}{\lambda}\sigma_{q_2p_2} \\ \\
\sigma_{p_1q_1} & \sigma_{p_1q_2} & \sigma_{p_1p_1} &
\frac{1}{\lambda}\sigma_{p_1p_2} \\
\\\frac{1}{\lambda}\sigma_{p_2q_1} &
\frac{1}{\lambda}\sigma_{p_2q_2} & \frac{1}{\lambda}\sigma_{p_2p_1}
& \frac{1}{\lambda^2}\sigma_{p_2p_2}
\end{array}\right). \end{equation}

We obtain a test of separability based on the
Robertson-Schr$\mathrm{\ddot o}$dinger uncertainty relation for the
partially scaled matrix $\sigma_\lambda$
\begin{equation}\det \left( \sigma_\lambda + \frac{i}{2}\Omega \right) \geq 0. \end{equation}

\section{Scaling Transform in the Three-Mode Case}

Changing the sign for one momentum only may not be efficient in the
case of three and more modes. The Peres-Horodecki criterion detects
the bipartite entanglement. This limitation can be overcome using
the scaling transform by changing momenta for several modes in
different ways. In the three-mode case, we can apply the following
transform:
\begin{equation}\left \{ {\begin{array}{lcl} p_1 \longrightarrow \lambda_1 \,p_1, \\ p_2 \longrightarrow \lambda_2 \, p_2, \\  p_3 \longrightarrow \lambda_3 \, p_3. \end{array}} \right. \end{equation}
The scaling parameters are determined, as in the previous case, in
the domain from $-1$ to $1$, namely, $\lambda_{1} \in [-1,1]$,
$\lambda_{2} \in [-1,1]$, and $\lambda_{3} \in [-1,1]$. The applied
transform leads to the scaled variance matrix
\begin{equation}\sigma_{\lambda_1 \lambda_2 \lambda_3} = \left(\begin{array}{cccccc}
\sigma_{q_1q_1} & \sigma_{q_1q_2} & \sigma_{q_1q_3} &
\frac{1}{\lambda_1}\sigma_{q_1p_1} &
\frac{1}{\lambda_2}\sigma_{q_1p_2} &
\frac{1}{\lambda_3}\sigma_{q_1p_3}
\\ \\
\sigma_{q_2q_1} & \sigma_{q_2q_2} & \sigma_{q_2q_3} &
\frac{1}{\lambda_1}\sigma_{q_2p_1} &
\frac{1}{\lambda_2}\sigma_{q_2p_2} & \frac{1}{\lambda_3}\sigma_{q_2p_3} \\ \\
\sigma_{q_3q_1} & \sigma_{q_3q_2} & \sigma_{q_3q_3} &
\frac{1}{\lambda_1}\sigma_{q_3p_1} &
\frac{1}{\lambda_2}\sigma_{q_3p_2} &
\frac{1}{\lambda_3}\sigma_{q_3p_3} \\ \\
\frac{1}{\lambda_1}\sigma_{p_1q_1} &
\frac{1}{\lambda_1}\sigma_{p_1q_2} &
\frac{1}{\lambda_1}\sigma_{p_1q_3} & \frac{1}{\lambda_1
^2}\sigma_{p_1p_1} &
\frac{1}{\lambda_1\, \lambda_2}\sigma_{p_1p_2} & \frac{1}{\lambda_1\, \lambda_3}\sigma_{p_1p_3}\\
\\ \frac{1}{\lambda_2}\sigma_{p_2q_1} &
\frac{1}{\lambda_2}\sigma_{p_2q_2} &
\frac{1}{\lambda_2}\sigma_{p_2q_3} & \frac{1}{\lambda_1 \,
\lambda_2}\sigma_{p_2p_1} & \frac{1}{\lambda_2^2}\sigma_{p_2p_2} &
\frac{1}{\lambda_2 \, \lambda_3}\sigma_{p_2p_3}
\\
\\ \frac{1}{\lambda_3}\sigma_{p_3q_1} &
\frac{1}{\lambda_3}\sigma_{p_3q_2} &
\frac{1}{\lambda_3}\sigma_{p_3q_3} & \frac{1}{\lambda_1 \,
\lambda_3}\sigma_{p_3p_1} & \frac{1}{\lambda_2 \,
\lambda_3}\sigma_{p_3p_2} & \frac{1}{\lambda_3^2}\sigma_{p_3p_3}
\end{array}\right). \end{equation}

The uncertainty relation for the scaled matrix demonstrates the
criterion of separability. Since the first three principal minors
remain unchanged, taking into consideration the initial uncertainty
relation, the criterion reduces to the three inequalities:
\begin{equation}\Sigma(\lambda_1, \lambda_2, \lambda_3) = \det \left[ \sigma_{\lambda_1 \lambda_2 \lambda_3} + \frac{i}{2}
\left(\begin{array}{cc}
0 & -I \\
I & 0
\end{array}\right) \right] \geq 0, \end{equation}
\begin{equation}\Sigma_5 (\lambda_1, \lambda_2) = \det \mathbf M_5 \left[\sigma_{\lambda_1 \lambda_2 \lambda_3} + \frac{i}{2}
\left(\begin{array}{cc}
0 & -I \\
I & 0
\end{array}\right) \right] \geq 0, \end{equation}
and
\begin{equation}\Sigma_3 (\lambda_1) = \det \mathbf M_4 \left[\sigma_{\lambda_1 \lambda_2 \lambda_3} + \frac{i}{2}
\left(\begin{array}{cc}
0 & -I \\
I & 0
\end{array}\right) \right] \geq 0. \end{equation}
Here $I$ is the $3 \times 3$ identity matrix. The fifth and fourth
principal minors are denoted as $\mathbf M_5$ and $\mathbf M_4$,
respectively. The criterion consists in checking the sign of
functions $\Sigma(\lambda_1, \lambda_2, \lambda_3)$, $\Sigma_5
(\lambda_1, \lambda_2)$, and $\Sigma_4 (\lambda_1)$ in the domain
$\lambda_1 \times \lambda_2 \times \lambda_3 = [-1, 1] \times [-1,
1] \times [-1, 1]$. The negativity of one function is an indicator
of entanglement of the given state with a certain variance matrix.

Below it will be shown that the absolute value of negative function
$\Sigma(\lambda_1, \lambda_2, \lambda_3)$ is connected with the
degree of entanglement.

We consider a specific pure three-mode Gaussian state with the wave
function as follows:
\begin{equation}\Psi(x,y,z) = N \exp \left( -\frac{x^2}{2}
-\frac{y^2}{2} -\frac{z^2}{2} + c_1 xy+ c_2 xz+ c_3 yz \right),
\end{equation}
where $c_1$, $c_2$, and $c_3$ are the parameters of quadratic form
determining the degree of entanglement. The normalization constant
$N$ reads
\begin{equation}N=\frac{\sqrt[4]{1-c_1 ^2 - c_2 ^2 - c_3 ^2 - 2c_1 c_2 c_3}}{\pi^{3/4}}. \end{equation}
In this case, the variance matrix $\sigma_{r_i r_j}$ can be
calculated using the definition (10). After calculating the elements
of $\sigma$, we obtain the following result:
\begin{equation}\sigma =
\frac{1}{2} \left( \begin{array}{cc}
U & 0 \\
0 & V
\end{array} \right),\end{equation}
where we use the notation
\begin{equation}\begin{array} {lcl} U=\left(1-c_1 ^2 - c_2 ^2 - c_3 ^2 - 2c_1 c_2
c_3 \right)^{-1} \left(\begin{array}{ccc}
1-c_3 ^2 & c_1 + c_2 c_3 &  c_2 + c_1 c_3\\
c_1 + c_2 c_3 & 1-c_2 ^2 & c_3 + c_1 c_2\\
c_2 + c_1 c_3 & c_3 + c_1 c_2 & 1-c_1 ^2
\end{array}\right),
\\
\\
V=\left(\begin{array}{ccc}
1 & -c_1 & -c_2 \\
-c_1 & 1 & -c_3 \\
-c_2 & -c_3 & 1
\end{array}\right).
\end{array} \end{equation}

The uncertainty relations for the state under study and the
normalization conditions determine the admissible domain of the
coefficients of the quadratic form. The following relations show the
possible domain of the coefficients:
\begin{equation}\left \{ {\begin{array}{lcl} c_1 \in (-1, 1), \\ c_2 \in (-1, 1), \\ c_3 \in (-1, 1), \\ 1-c_1 ^2 - c_2 ^2 - c_3 ^2 - 2c_1 c_2 c_3 >0. \end{array}} \right. \end{equation}

\begin{figure}[htbp]
  \begin{minipage}[h]{0.5\linewidth}
        \centering
        \includegraphics[width=211 pt]{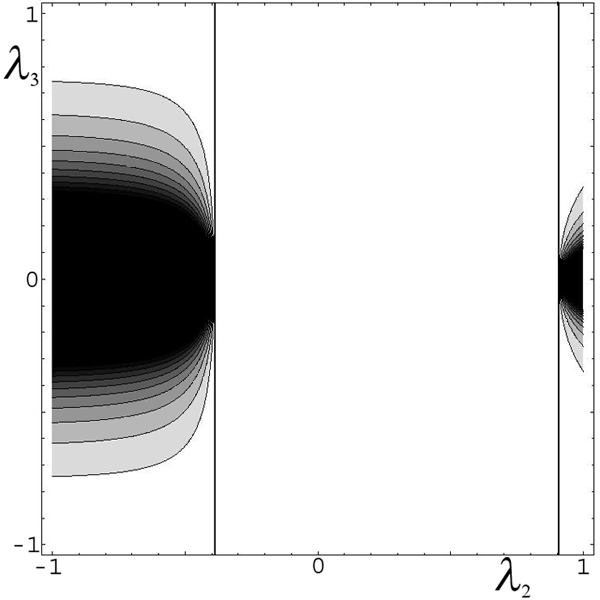}
        \end{minipage}
 \begin{minipage}[h]{0.5\linewidth}
        \centering
        \includegraphics[width=213 pt]{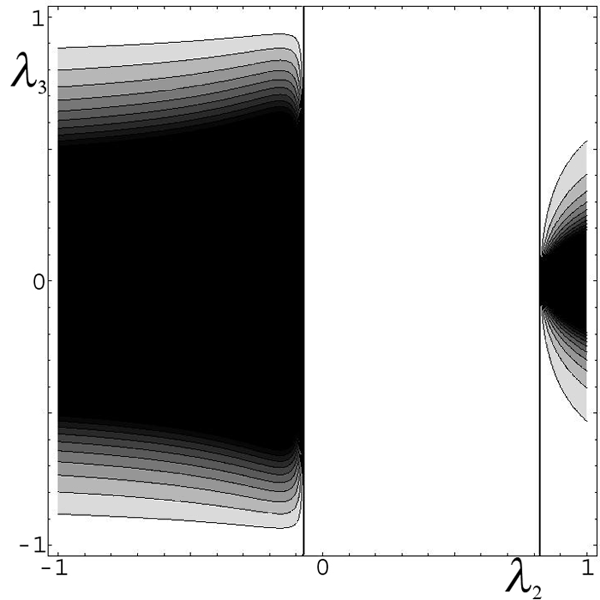}
        \end{minipage}
  \flushleft \hangindent=0.55cm \hangafter=0 \small \textbf{Fig. 1.} Contour plots of ${\Sigma(\lambda_1, \lambda_2, \lambda_3)}$ for the wave function (36), where $\lambda_1 = 1/2$ and $c_1 = 2/3$ (on the left) and $c_1 =
  5/6$ (on the right).
\end{figure}

\begin{figure}[t]
  \begin{minipage}[h]{0.5\linewidth}
        \centering
        \includegraphics[width=210 pt]{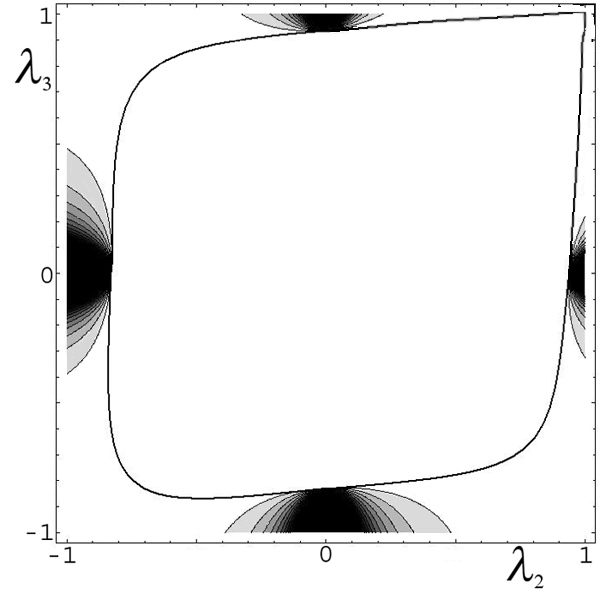}
 \end{minipage}
 \begin{minipage}[h]{0.5\linewidth}
        \centering
        \includegraphics[width=210 pt]{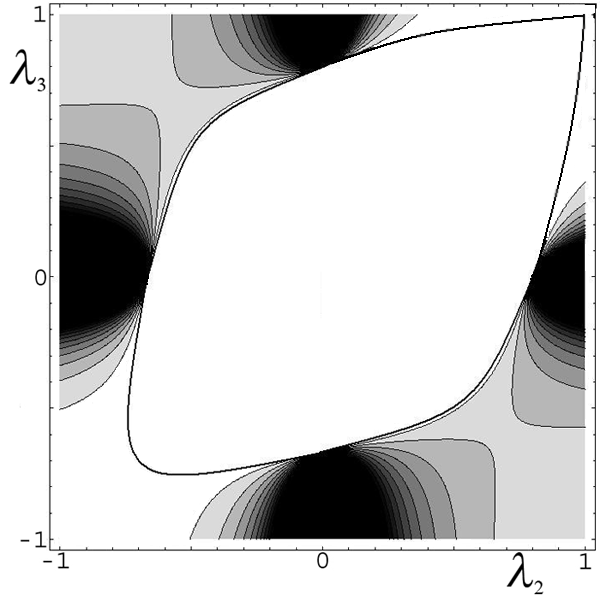}
        \end{minipage}
  \flushleft \hangindent=0.55cm \hangafter=0 \small \textbf{Fig. 2.} Contour plots of ${\Sigma(\lambda_1, \lambda_2, \lambda_3)}$ for $c_1 = c_2 = 1/4$, where $\lambda_1 = 1/2$ and $c_3 = 1/4$ (on the left) and $c_3 = 1/2$ (on the right).
\end{figure}

The Peres-Horodecki criterion for the three-mode case is obtained by
applying the transform of quadratures as follows:
\begin{equation}\left \{ {\begin{array}{lcl} p_1 \longrightarrow p_1, \\ p_2 \longrightarrow -p_2, \\  p_3 \longrightarrow -p_3. \end{array}} \right. \end{equation}
The Robertson-Schr$\mathrm{\ddot o}$dinger uncertainty relation for
the transformed matrix provides two inequalities - one from
determinants of the whole variance matrix and the other from the
fifth principal minor of the matrix. The determinant [analogous to
(24) in the three-mode case] is equal to zero for any $c_1$, $c_2$,
and $c_3$. The determinant of the fifth principal minor is
proportional to $c_1 ^2$, which leads to the conclusion that the
Peres-Horodecki criterion does not detect any entanglement for all
wave functions of the form (30) with $c_1 = 0$.

Below we examine the application of the partial scaling criterion to
the wave function (30) for certain coefficients $c_1$, $c_2$, and
$c_3$.

First, we consider the case where the entanglement is given only by
the term $xy$ or, in the language of coefficients, the case ${c_2 =
c_3 =0}$. Obviously, the problem is symmetric with respect to
permutations of $c_1$, $c_2$ and $c_3$. From relations (30) and (31)
and conditions (34), we have the following form of wave function for
any $c_1 \in (-1, 1)$
\begin{equation}\Psi(x,y,z) = \frac{\sqrt[4]{1-c_1 ^2}}{\pi^{3/4}} \exp \left( -\frac{x^2}{2} -\frac{y^2}{2} -\frac{z^2}{2} + c_1 xy \right). \end{equation}
Simplifying the variance matrix (32) and applying the partial
scaling transform (25) we obtain
\begin{equation}\Sigma(\lambda_1, \lambda_2,
\lambda_3) = -\frac{1}{64\lambda_1 ^2 \lambda_2 ^2 \lambda_3 ^2}
\frac{c_1 ^2 (1-\lambda_1 \lambda_2)^2 - (1-\lambda_1
^2)(1-\lambda_2 ^2)}{1-c_1 ^2}.
\end{equation}

\begin{wrapfigure}{O}{80mm}
    \centering
    \includegraphics[width=70mm,height=67mm]{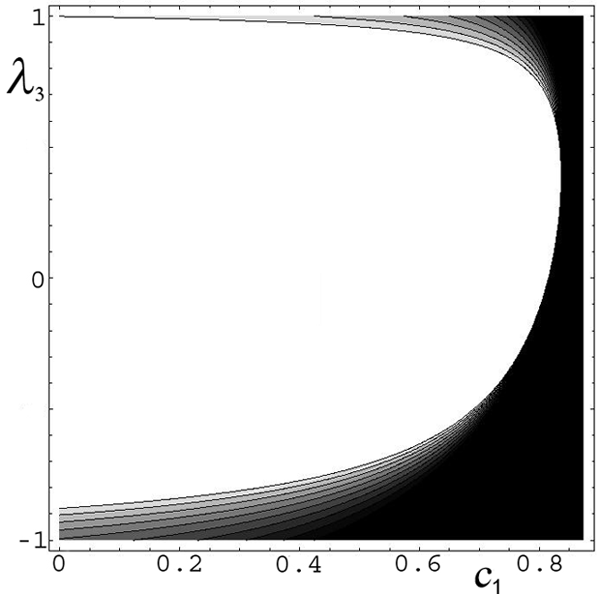}
\flushleft \hangindent=0.7cm \hangafter=0 \small \textbf{Fig. 3.}
Contour plot of ${\Sigma(\lambda_1, \lambda_2, \lambda_3)}$ for $c_2
= c_3 = 1/4$, where $\lambda_1 = 1/2$, $\lambda_2 = 1/4$.
\end{wrapfigure}

For $c_1 =0$, expression (37) is nonnegative in the domain
$\lambda_1 \times \lambda_2 \times \lambda_3 = [-1, 1] \times [-1,
1] \times [-1, 1]$. So, for this separable state, the uncertainty
relation holds and the criterion does not show any entanglement.
Otherwise, for any $c_1 \in (-1, 1)$, such a pair $\lambda_1$ and
$\lambda_2$ exists for which $\Sigma(\lambda_1, \lambda_2,
\lambda_3)$ is negative. Thus, the partial scaling criterion (27)
detects the entanglement in this case. Moreover, it is easy to see
that the rate and area of negativity increases with the coefficient
$c_1$. These facts will be demonstrated by some examples below.

An illustrative example is shown in Fig. 1. Contour plots present
the function ${\Sigma(\lambda_1, \lambda_2, \lambda_3)}$ with
$\lambda_1 = 1/2$. To show that the graph is moving down with
increase in entanglement, we constructed contour plots for $c_1 =
2/3$ and $5/6$. The negative area of the graph for bigger
coefficient $c_1$ is larger and lower than for smaller $c_1$. Since
we are not interested in the positive part of the graphs, the grey
scale shows only the rate of negativity. The negative part of the
graphs is bounded by lines, where the function is equal to zero. The
area between lines is a positive part of the function. It is worth
mentioning that for ${c_1 \rightarrow 1}$ or, in other words, for
the maximum entanglement in this case, the positive part collapses
to the line $\lambda_2 = \lambda_1$. In the considered case, it is
the line $\lambda_2 = 1/2$.

The completely entangled state leads to the same result. We
demonstrate it by the case $c_1 =c_2 =1/4$. In Fig. 2 graphs for
$c_3 = 1/4$ and $1/2$ are shown. Obviously, the graph for bigger
$c_3$ has larger negative area and its absolute value in points,
where both functions are negative, is bigger. The zero level is
shown by a black curve. The area inside it contains only positive
values.

A vivid illustration that the graph is mowing down with increase in
entanglement is shown in Fig. 3 where the function is constructed
for the wave function with $c_2 =c_3 = 1/4$. As before, it
represents ${\Sigma(\lambda_1, \lambda_2, \lambda_3)}$ with
$\lambda_1 = 1/2$ and $\lambda_2 = 1/4$. Obviously, the positive
part of plot (white color) diminishes with increase in $c_1$ from
$0$ to $7/8$ and the negative part is moving down.

\section{Scaling Transform in the Four-Mode Case}

To apply the graphing method discussed in the previous sections to
the four-mode state, we apply the transform of momenta of the form
\begin{equation}\left \{ {\begin{array}{lcl} p_1 \longrightarrow \lambda_1 \,p_1, \\ p_2 \longrightarrow \lambda_2 \, p_2, \\  p_3 \longrightarrow \lambda_3 \, p_3, \\ p_4 \longrightarrow \lambda_4 \, p_4. \end{array}} \right. \end{equation}

\begin{figure}[htbp]
  \begin{minipage}[h]{0.5\linewidth}
        \centering
        \includegraphics[width=211 pt]{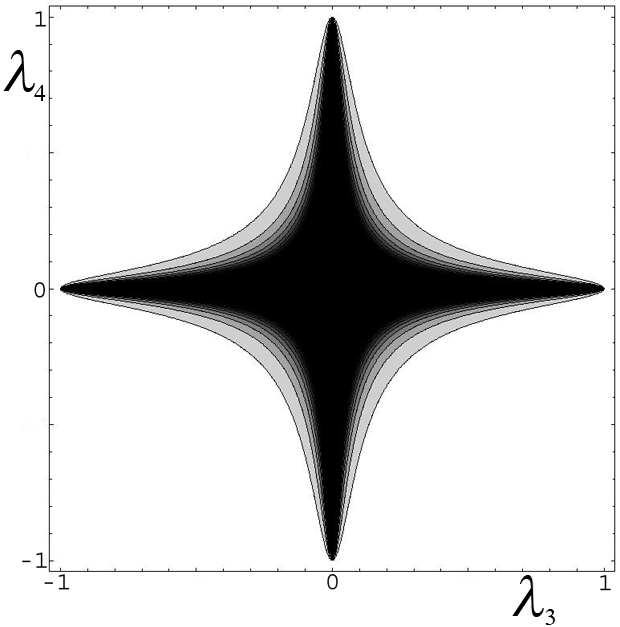}
   \end{minipage}
 \begin{minipage}[h]{0.5\linewidth}
        \centering
        \includegraphics[width=213 pt]{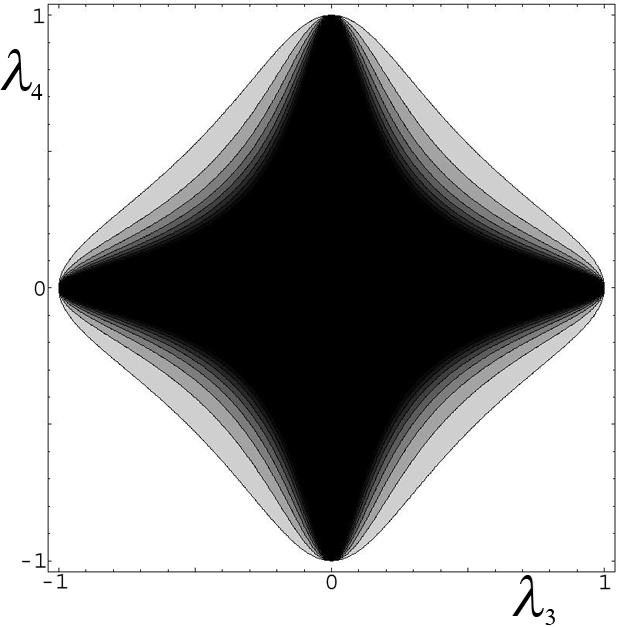}
     \end{minipage}
  \flushleft \hangindent=0.55cm \hangafter=0 \small \textbf{Fig. 4.} Contour plots of ${\Sigma(\lambda_1, \lambda_2, \lambda_3, \lambda_4)}$ for the wave function (40), where $\lambda_1 = -1$ and $\lambda_2 = 1/2$ and $c_1 = 1/8$ (on the left) and $c_1 =
  1/2$ (on the right).
\end{figure}

All scaling parameters belong to the interval $[-1,1]$. The
transformed variance $8 \times 8$ matrix $\sigma_{\lambda_1
\lambda_2 \lambda_3 \lambda_4}$ is similar to (26). The
Robertson-Schr$\mathrm{\ddot o}$dinger uncertainty relation can be
easily reduced to four inequalities, in complete analogy to the
previous case. We consider only the inequality for determinant of
this matrix. It will be shown that the function defined below gives
results very similar to the ones obtained in the three-mode case.
The inequality under consideration reads
\begin{equation}\Sigma(\lambda_1, \lambda_2, \lambda_3, \lambda_4) = \det \left[ \sigma_{\lambda_1 \lambda_2 \lambda_3 \lambda_4} + \frac{i}{2}
\left(\begin{array}{cc}
0 & -I \\
I & 0
\end{array}\right) \right] \geq 0. \end{equation}

It is worth mentioning that the Peres-Horodecki criterion applied to
the four-mode wave function does not give complete information on
entanglement -- in some cases, it does not work.

We investigate wave functions of the form analogous to the ones
considered in the three-mode case. First, let us look at the
function
\begin{equation}\Psi(x,y,z,k)
= N \exp \left( -\frac{x^2}{2} -\frac{y^2}{2} -\frac{z^2}{2}
-\frac{k^2}{2} + c_1 xy \right), \end{equation} where
\begin{equation}
N=\frac{\sqrt[4]{1-c_1 ^2}}{\pi}.\end{equation}

Graphs of $\Sigma(-1,1/2,\lambda_3, \lambda_4)$ for $c_1 = 1/8$ and
$1/2$ are shown in Fig. 4. One can observe the same effect of moving
the graph down with increase in entanglement. Entire graphs are
located in the nonpositive half-space.

For completeness, in Fig. 5 contour plots for completely entangled
states are shown. The graphs are constructed for
\begin{equation}\Psi(x,y,z,k) = N \exp \left( -\frac{x^2}{2} -\frac{y^2}{2} -\frac{z^2}{2} -\frac{k^2}{2} + \frac{1}{4}xy + \frac{1}{4}xz + \frac{1}{4}xk + \frac{1}{4}yz + \frac{1}{4}yk + c_6 zk \right), \end{equation}
where $c_6 = 1/8$ and $1/2$.

\begin{figure}[t]
  \begin{minipage}[h]{0.5\linewidth}
        \centering
        \includegraphics[width=209 pt]{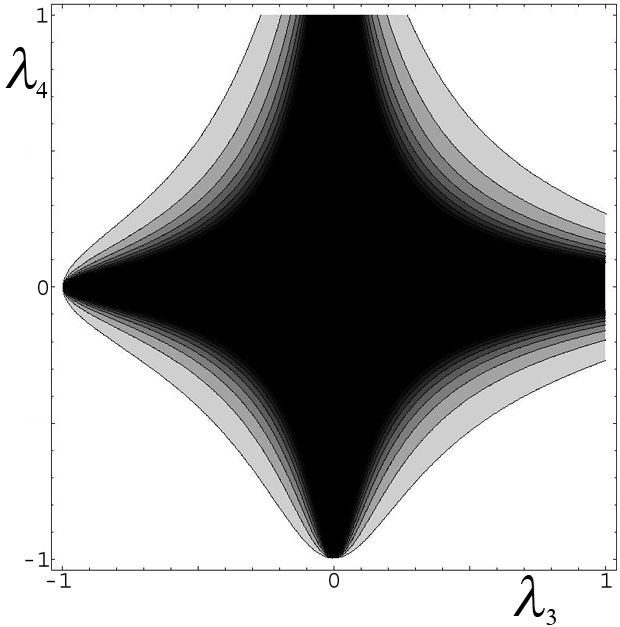}
        \end{minipage}
 \begin{minipage}[h]{0.5\linewidth}
        \centering
        \includegraphics[width=213 pt]{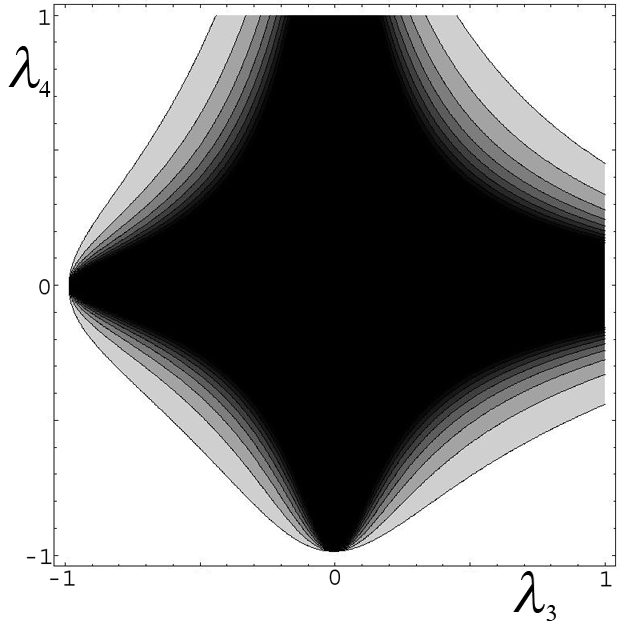}
        \end{minipage}
  \flushleft \hangindent=0.55cm \hangafter=0 \small \textbf{Fig. 5.} Contour plots of ${\Sigma(\lambda_1, \lambda_2, \lambda_3, \lambda_4)}$ for the wave function (42), where $\lambda_1 = -1$ and $\lambda_2 = 1/2$ and $c_6 = 1/8$ (on the left) and $c_6 = 1/2$ (on the right).
\end{figure}

In both cases, the partial scaling transform detects the
entanglement. Also it can be seen that the bigger the entanglement,
the lower the location of the plot of $\Sigma(\lambda_1, \lambda_2,
\lambda_3, \lambda_4)$. The two functions are nonpositive in the
domain $\lambda_3 \times \lambda_4 = [-1,1]\times [-1,1]$ too.

\section{Scaling Transform Criterion for Mixed Gaussian States}

In this section, the partial scaling method is tested on the
applicability for mixed Gaussian states. First, we consider the
three-mode state with the variance matrix
\begin{equation}\sigma = \left(\begin{array}{cccccc}
\frac{6}{5} & \frac{1}{5} & \frac{1}{5} & \frac{1}{10} &
\frac{1}{10} & \frac{1}{10}
\\ \\
\frac{1}{5} & \frac{6}{5}  & \frac{1}{5} & \frac{1}{10} &
\frac{1}{10} & \frac{1}{10} \\ \\
\frac{1}{5} & \frac{1}{5} & \frac{6}{5} & \frac{1}{10} &
\frac{1}{10} &
\frac{1}{10} \\ \\
\frac{1}{10} & \frac{1}{10} & \frac{1}{10} & \frac{1}{2} &
-\frac{1}{8} & -\frac{1}{8}\\
\\ \frac{1}{10} &
\frac{1}{10} & \frac{1}{10} & -\frac{1}{8} & \frac{1}{2} &
-\frac{1}{8}
\\
\\ \frac{1}{10} &
\frac{1}{10} & \frac{1}{10} & -\frac{1}{8} & -\frac{1}{8} &
\frac{1}{2}
\end{array}\right). \end{equation}

The graph for $\Sigma(1,\lambda_2 , \lambda_3)$ is presented in Fig.
6 (on the left) where the positive part of the graph is shown with
white color. Obviously, the partial scaling method detects the
entanglement in this case -- the graph has a negative area in the
domain $\lambda_2 \times \lambda_3 = [-1,1] \times [-1,1]$.

\begin{figure}[h]
  \begin{minipage}[h]{0.5\linewidth}
        \centering
        \includegraphics[width=207 pt]{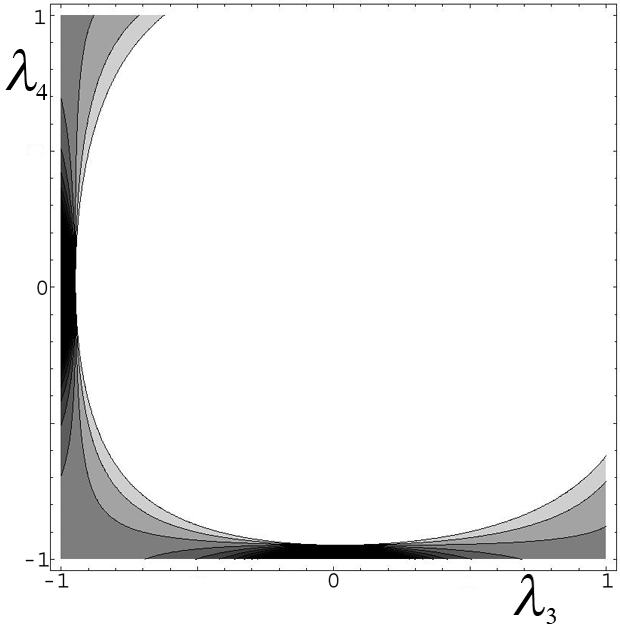}
   \end{minipage}
 \begin{minipage}[h]{0.5\linewidth}
        \centering
        \includegraphics[width=211 pt]{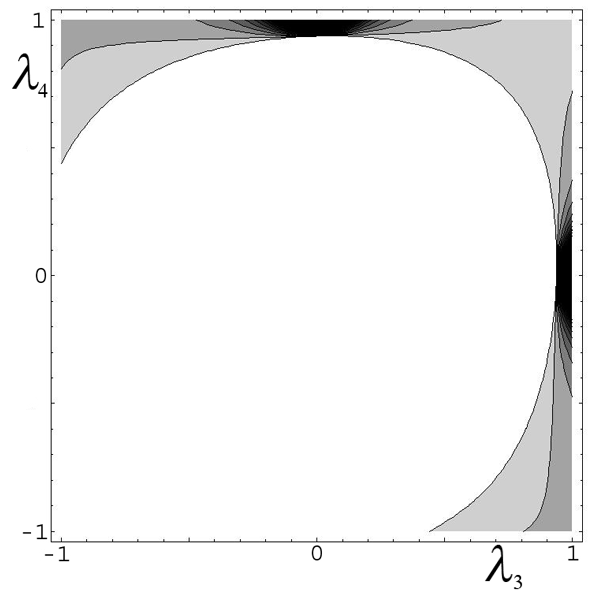}
     \end{minipage}
  \flushleft \hangindent=0.55cm \hangafter=0 \small \textbf{Fig. 6.} Contour plot of ${\Sigma(\lambda_1, \lambda_2, \lambda_3)}$ for the variance matrix (43), where $\lambda_1 = 1$ (on the left) and contour plot of ${\Sigma(\lambda_1, \lambda_2, \lambda_3, \lambda_4)}$ for the variance matrix (44), where $\lambda_1 = -1$ and $\lambda_2 = 1/2$ (on the right).
\end{figure}

A similar result is obtained for the four-mode mixed Gaussian state.
In Fig. 6 (on the right) the graph for $\Sigma(-1,1/2,\lambda_3 ,
\lambda_4)$ is shown.

The $8 \times 8$ variance matrix reads
\begin{equation}\sigma = \left(\begin{array}{cccccccc}
\frac{8}{5} & \frac{2}{5} & \frac{2}{5}  & \frac{2}{5} &
\frac{1}{10} & \frac{1}{10} & \frac{1}{10} & \frac{1}{10}
\\ \\
\frac{2}{5} & \frac{8}{5}  & \frac{2}{5} & \frac{2}{5} &
\frac{1}{10} &
\frac{1}{10} & \frac{1}{10} & \frac{1}{10} \\ \\
\frac{2}{5} & \frac{2}{5} & \frac{8}{5} & \frac{2}{5} & \frac{1}{10}
& \frac{1}{10} &
\frac{1}{10} & \frac{1}{10}
\\ \\
\frac{2}{5} & \frac{2}{5} & \frac{2}{5} & \frac{8}{5} & \frac{1}{10}
& \frac{1}{10} &
\frac{1}{10} & \frac{1}{10} \\ \\
\frac{1}{10} & \frac{1}{10} & \frac{1}{10} & \frac{1}{10} &
\frac{1}{2} & -\frac{1}{8} & -\frac{1}{8} &
-\frac{1}{8}\\
\\ \frac{1}{10} &
\frac{1}{10} & \frac{1}{10} & \frac{1}{10} & -\frac{1}{8} &
\frac{1}{2} & -\frac{1}{8}  & -\frac{1}{8}
\\ \\ \frac{1}{10} &
\frac{1}{10} & \frac{1}{10} & \frac{1}{10} & -\frac{1}{8} &
-\frac{1}{8} & \frac{1}{2} & -\frac{1}{8}
\\ \\ \frac{1}{10} &
\frac{1}{10} & \frac{1}{10} & \frac{1}{10} & -\frac{1}{8} &
-\frac{1}{8} & -\frac{1}{8} & \frac{1}{2}
\end{array}\right). \end{equation}

In both plots, there is a negative area, which depicts the
entanglement of states with the variance matrices (43) and (44).

\section{Conclusions}

To conclude, we point out the main results of our work.

We have shown that the new criterion of entanglement based on the
scaling transform of the photon quadrature components in the Wigner
function of the quantum state provides the possibility to detect the
entanglement for three- and four-mode fields. The calculated
principal minors of the scaled quadrature dispersion matrices
shifted by the quadrature commutator matrix become negative for
entangled Gaussian states. The values of modulus of the negative
principal minors can serve as empirical characteristics of the
degree of entanglement. The results obtained for the three-mode
entangled Gaussian states, in general, agree with the previous
results of \cite{Chirkin1, Chirkin2} where the scaling criterion of
entanglement was applied to the photon states created due to the
light interaction with nonlinear crystal. In a future publication,
we will extend the analysis of the entanglement based on the scaling
criterion to the multimode light beams (both of Gaussian and
non-Gaussian tipes) and compare this criterion with other approaches
to the problem of detecting the entanglement.

\section{Acknowledgments}
V.I.M. acknowledges the support of the Russian Foundation for Basic
Research under Project No. 07-02-00598.

\end{document}